\documentstyle[12pt,epsfig,rotating]{article}
\begin{document}

\begin{titlepage}
\begin{center}
\large STATE RESEARCH CENTER OF RUSSIA\\
\Large INSTITUTE FOR HIGH ENERGY PHYSICS\\
\end{center}

\bigskip

\begin{flushright}
{Preprint IHEP}
{ 98-62\\}
\end{flushright}     

\bigskip

\begin{center}{\Large STUDY OF ${\bf\eta\pi^-}$ PRODUCTION 
BY PIONS IN THE COULOMB FIELD }
\end{center}

\bigskip

  \begin{center}
D.V.~Amelin, E.B.~Berdnikov, S.I.~Bityukov,\\
G.V.~Borisov, R.I.~Dzhelyadin, V.A.~Dorofeev, \\
A.V.~Ekimov, Yu.P.~Gouz$^1$, A.K.~Konoplyannikov, \\
A.N.~Karyukhin, I.A.~Kachaev, Yu.A.~Khokhlov, \\
V.F.~Konstantinov, S.V.~Kopikov, M.E.~Kostrikov, \\
V.V.~Kostyukhin, S.A.~Likhoded, V.D.~Matveyev, \\
A.P.~Ostankov, D.I.~Ryabchikov, O.V.~Solovianov, \\
A.A.~Solodkov, E.A.~Starchenko, \fbox{N.K.~Vishnevsky}$\:$, \\
E.V.~Vlassov, A.M.~Zaitsev$^2$ \\
IHEP, Protvino, Russia\\
G.G.~Sekhniaidze, E.G.~Tskhadadze,\\
IPh, Tbilisi, Georgia
\end {center}

\vspace{2cm}
  
\begin{abstract}
\noindent
The production of $\eta\pi^-$ system at low $M_{\eta\pi^-}$ by the $\pi^-$ 
beam in the Coulomb field of Be nuclei was studied. 
The cross section of the reaction \mbox{$\pi^-Be\rightarrow\eta\pi^-Be$} 
was compared to the experimental data on the decay 
\mbox{$\eta \rightarrow \pi^+ \pi^- \gamma$} 
and to the theoretical predictions.
\end{abstract}

\vspace{1cm}

\begin{flushright}
Submitted to {\it Yadernaja Fizika}
\end{flushright}

\vspace{1cm}

\small
\noindent
\rule{3cm}{0.5pt}\\
$^1$E-mails: gouz@mx.ihep.su,~~Iouri.Gouz@cern.ch\\
$^2$E-mails: zaitsev@mx.ihep.su,~~Alexandre.Zaitsev@cern.ch\\

\vspace{1cm}

\begin{center}
Protvino 1998
\end{center}

\end{titlepage}

\newpage
\section*{Introduction.}
\begin{figure}[ht]
\epsfxsize=\textwidth
\epsfysize=0.7\textwidth
\epsffile{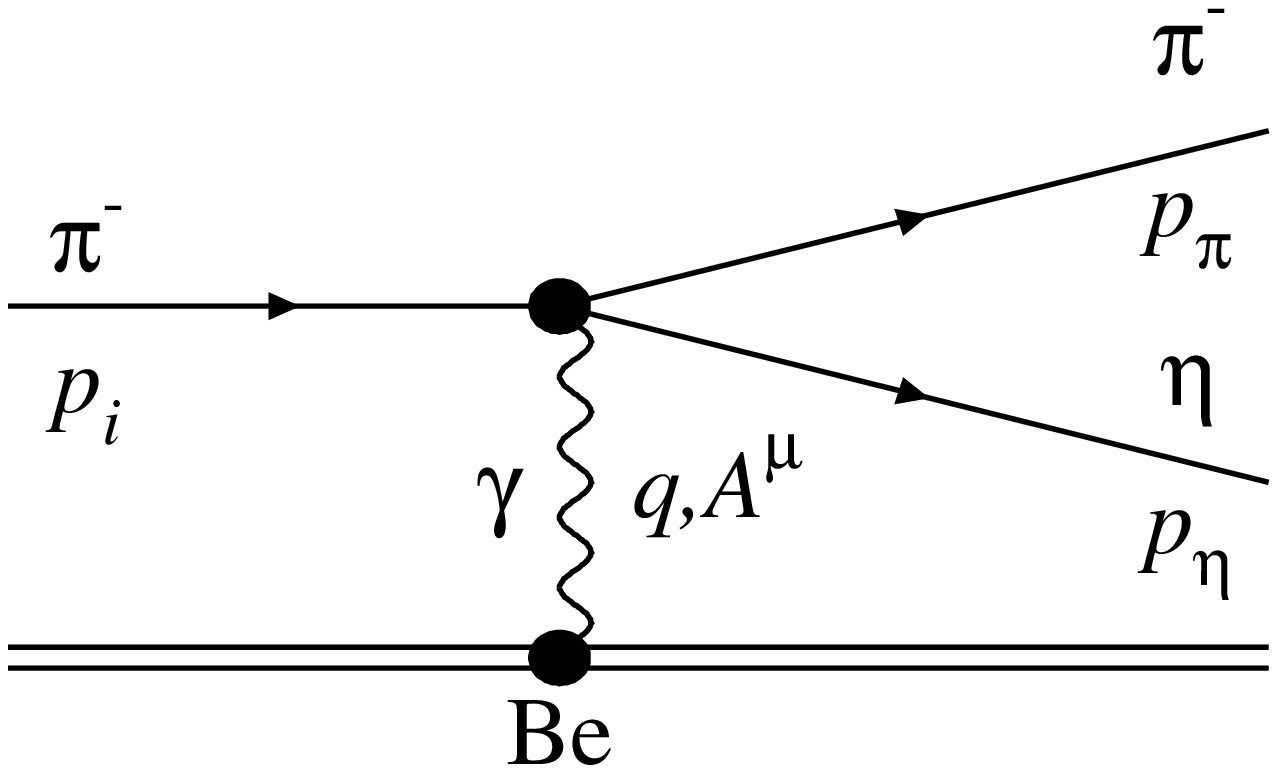}
\begin{picture}(160,0)
\end{picture}
\caption{Reaction (\protect\ref{prima})} 
\label{primdiag}
\end{figure}

\indent     
We present here the study of the $\eta \pi^-$ system production 
at low mass ($M_{\eta\pi^-} < 1.18 \mbox{ GeV}/c^2$)
in the Coulomb field of $Be$ nuclei
\begin{equation}
\pi^- Be \rightarrow \eta \pi^- Be.
\label{prima}
\end{equation}
The subject of our study is the process
\begin{equation}
\pi^- \gamma \rightarrow \eta \pi^-. 
\label{box}
\end{equation}  
Its amplitude at low $M_{\eta\pi^-}$ can be expressed in the following form:
\begin{equation} 
M_{\eta \pi \pi \gamma} = 
    -i \epsilon_{\mu \nu \rho \sigma} A^{\mu} p_{i}^{\nu} p_{\pi}^{\rho} 
    p_{\eta}^{\sigma} F_{\eta \pi \pi \gamma}(s,t,u)
\label{amplbox}
\end{equation}
(notations are given in Fig.\ref{primdiag}, $s=(p_{\pi}+p_{\eta})^2$, 
$t=(p_{i}-p_{\pi})^2$, $u=(p_{i}-p_{\eta})^2$).

In the chiral limit the amplitude of the process (\ref{box})
is determined by the box anomaly, and $F_{\eta\pi\pi\gamma}(0,0,0)$ 
can be expressed as follows \cite{ko,holstein}
\cite{ko,holstein} 
\begin{equation}
F_{\eta \pi \pi \gamma} (0,0,0) = \frac{e}{4 \pi^2 f_{\pi}^3} 
(\frac{f_{\pi}}{f_8} \frac{cos \theta _p}{\sqrt{3}} - 
\frac{f_{\pi}}{f_0} \sqrt{\frac{2}{3}} sin \theta_p ). 
\label{amplko}
\end{equation}
Here $f_{\pi,0,8}$ are the pion, singlet and octet decay constants, 
$\theta_p$ is the singlet-octet mixing angle for pseudoscalars.

The value of $F_{\eta\pi\pi\gamma}$ at the chiral limit, 
as well as its dependence on the kinematical variables, 
are of interest for the theoretical analysis. 

A model predicting the dependence of $F_{\eta\pi\pi\gamma}(s,t,u)$ 
on the kinematical variables was described in \cite{holstein}; 
following this model, the authors performed the fit of experimental data 
on radiative decays of light pseudoscalar mesons
\begin{eqnarray}
\label{ppgdecay}
\eta,\eta' & \rightarrow & \pi^+\pi^-\gamma ,\\
\label{ggdecay}
\eta,\eta' & \rightarrow & \gamma\gamma
\end{eqnarray}
and determined $f_0$, $f_8$ and $\theta_p$. For $F_{\eta\pi\pi\gamma}$ 
this fit yields
\begin{equation}
F_{\eta\pi\pi\gamma}(0,0,0)=6.5 \pm 0.3 \mbox{ GeV}^{-3}.
\label{fsimple}
\end{equation}

The decays (\ref{ppgdecay}) proceed at positive $t=M_{\pi\pi}^2$. 
Study of the reaction (\ref{box}) gives information on 
$F_{\eta\pi\pi\gamma}(s,t,u)$ in different region of kinematical variables, 
at negative $t$. Here we compare the experimentally measured cross section 
of the reaction (\ref{prima}) at low $M_{\eta\pi^-}$ with predictions 
of \cite{holstein}.

\section{Main features of the $\eta\pi^-$ system production 
in the Coulomb field of nuclei}
     
The cross section of the $\eta\pi^-$ system production in the Coulomb 
field of nuclei is given by the expression \cite{primakoff}
\begin{equation} 
\frac{ d \sigma}{ ds dq^2} = 
\frac{Z^2 \alpha}{ \pi} \frac{|q^2-q_{min}^2|}{q^4} 
\frac{1}{s-m_{\pi}^2} \frac{p_{\eta}^3 p_i}{48 \pi} 
F_{\eta \pi \pi \gamma }^2  G^2(q^2).
\label{xprim}
\end{equation}

Here $Z$ is the charge of nucleus, $G(q^2)$ is its electromagnetic 
form-factor, $p_i$ and $p_{\eta}$ are absolute values of momenta 
of incident pion and outgoing $\eta$-meson in the c.m. system 
of produced $\eta$ and $\pi^-$.

The differential cross section of this reaction is strongly peaking 
at low $q^2$. 
This feature facilitates the event selection; 
from the other hand, it leads to the fast drop of the cross section 
at higher masses of produced $\eta\pi^-$ due to the cut 
on $|q_{min}^2| \simeq \frac{(s-m_{\pi}^2)^2}{4E_{beam}^2}$.

In case of $\eta\pi^-$ production at low mass in the Coulomb field, 
the $P$-wave is predominantly produced\footnote{
Note that the $P$-wave in the $\eta\pi^-$ system has an exotic set 
of quantum numbers, $I^G J^P = 1^- 1^-$.}
($S$-wave corresponds to 0--0 transition and is suppressed 
$\propto q^2/m^2_{\rho}$, 
and higher waves are suppressed by the barrier factor $p^{2l}$). 
However higher waves appear in the amplitude (\ref{amplbox}) 
due to the dependence of $F_{\eta\pi\pi\gamma}$ on kinematical variables, 
and their fraction increases with $M_{\eta\pi^-}$.

The background for the process of production in Coulomb field 
is the process of hadronic production of the $\eta\pi^-$ system.
In the hadronic production of $\eta\pi^-$ at high energies 
the amplitudes with positive exchange naturality with projection of 1
onto the Gottfried-Jackson axis are dominating \cite{epbec}.
At low $|q^2|$ this is mainly diffractive processes, which proceed 
coherently on nuclei.
They have characteristic $q^2$ distribution in form
\begin{equation}
\frac{d\sigma}{dq^2} \sim |q^2-q^2_{min}| e^{- b |q^2|}
\label{xdiffr}
\end{equation}
with slope parameter $b\simeq 50\mbox{ GeV}^{-2}$
and distribution on azimuthal angle in Gottfried-Jackson 
frame (Treiman-Yang angle)
\begin{equation}
\frac{dN}{d\varphi_{\sc ty}} \sim \sin^2\varphi_{\sc ty}.
\label{phidif}
\end{equation}
The $\eta\pi^-$ can also be produced in processes with negative 
exchange naturality (e.g. by $b_1$-trajectory).
Such exchanges, however, are suppressed in scattering on nuclei at high
energies, which can be illustrated by the absence of signal from 
$a_0(980)$-meson in the $\eta\pi^-$ effective mass spectrum \cite{epbec}.
At low $M_{\eta\pi^-}$ for the contribution of such exchanges one can 
expect zero projection onto the Gottfried-Jackson axis
(uniform distribution on Treiman-Yang angle)
and broad $q^2$-distribution of type
\begin{equation}
\frac{d\sigma}{dq^2} \sim e^{-b|q^2|}.
\label{xnega}
\end{equation}
with slope parameter $b\simeq 7\mbox{ GeV}^{-2}$.
As the contribution of negative naturality exchanges 
is expected to be small, such approximate description
is sufficient for our studies.

Apart from hadronic production of $\eta\pi^-$, there exists 
instrumental background, which consists of events with 
different final state, where one or more slow particles 
were not detected.
For this background one can expect broad $q^2$-distribution,
which can also be parametrized in form 
$d\sigma/dq^2 \sim e^{-b|q^2|}$. 
  
\section{The VES setup}
     
The experiment is performed at the VES setup (IHEP, Protvino) 
with $\pi^-$ beam at the momentum of 37 GeV/{\it c}.
The setup is a large aperture magnetic spectrometer which includes a system 
of proportional and drift chambers and a lead-glass electromagnetic 
calorimeter. 
The target was beryllium ($l=4\mbox{ cm}$). 
The trigger conditions required presence of two or more charged particles 
in the forward hemisphere and absence of hard charged particles 
in the backward hemisphere. 
The process under study (\ref{prima}) was detected with $\eta$-meson 
decaying into $\pi^+ \pi^- \pi^0$.
In the region of $M_{\eta\pi^-}=1\mbox{ GeV}$, the setup resolution 
on the effective mass of $\eta\pi^-$ is 
$\sigma(M_{\eta\pi^-}) \simeq 15 \mbox{ MeV}$, 
and on the transverse momentum $\sigma(p_t) \simeq 17\mbox{ MeV}$.

\section{Event selection, results}
     
The main event selection criteria were the following:
\begin{itemize}     
\item the event contains three tracks of charged particles 
with total charge of -1 and two photons in the final state;
\item the total energy of the final state lies within the interval 
of $36<E_{tot}<39$~GeV;
\item the charged particles are not identified as electrons;
\item the effective mass of two photons lies within the interval 
of the $\pi^0$ mass: $105<m_{\gamma\gamma}<165$~MeV;
\item the effective mass of one of $\pi^+ \pi^- \pi^0$ subsystems 
lies in the interval of the $\eta$-meson mass: 
$531<m_{\pi^+ \pi^- \pi^0}<567$~MeV. 
The constraint $m_{\gamma\gamma}=m_{\pi^0}$ was applied for the calculation 
of $m_{\pi^+\pi^-\pi^0}$.
\end{itemize}     

Fig.\ref{otbor} shows $m_{\gamma\gamma}$ (a) and $m_{\pi^+\pi^-\pi^0}$ (b) 
spectra. 
Signals from $\pi^0$ and $\eta$ mesons are clearly seen over smooth 
background. 

Fig.\ref{m_etapi} shows the $M_{\eta\pi^-}$ spectrum. 
It is dominated by $a_2^-(1320)$-meson signal. 
The arrow shows the cut used for the selection of events near threshold: 
$M_{\eta\pi^-}<1.18\mbox{ GeV}$.
The background for this spectrum was evaluated and subtracted using the 
events from control intervals on $m_{\pi^+\pi^-\pi^0}$ 
around the $\eta$-meson peak, with $503<m_{\pi^+ \pi^- \pi^0}<521$~MeV 
or $577<m_{\pi^+ \pi^- \pi^0}<595$~MeV.
For all other spectra the method of bin filtering was used, 
i.e. the number of events in each bin was determined as the number 
of $\eta$-mesons obtained by fitting of the $m_{\pi^+\pi^-\pi^0}$ 
spectrum for this bin.

Fig.\ref{tcoul} shows the $|q^2-q^2_{min}|$ distribution for selected events. 
It has characteristic Coulomb peak at low $|q^2-q^2_{min}|$, 
as well as a broad structure corresponding to the hadronic production 
of $\eta\pi^-$.
The result of the fit by superposition of $q^2$-distributions for 
Coulomb (\ref{xprim}) and hadronic (\ref{xdiffr}) production of $\eta\pi^-$,
convoluted with experimental resolution of the setup, 
and instrumental background, is shown by dashed line in Fig.\ref{tcoul}.

The parameters of the instrumental background were determined by
the analysis of distributions on total energy and Treiman-Yang 
angle for the events with
$0.01<|q^2-q_{min}^2|<0.1\mbox{ GeV}^2$ and $M_{\eta\pi^-}<1.18\mbox{ GeV}/c^2$
(Fig. \ref{figfi}a,b; in the first case the aforementioned selection 
on total energy was not applied).
In the total energy distribution one can see the peak corresponding 
to fully reconstructed events.
The background to the left of the peak, with $34<E_{tot}<35.5\mbox{ GeV}$, 
has uniform distribution on Treiman-Yang angle and $q^2$ distribution 
with slope parameter $b\simeq 7\mbox{ GeV}^{-2}$.
The number of background events was determined by fitting of the 
distribution on Treiman-Yang angle by superposition of (\ref{phidif})
and constant. 
It is worth noticing that similar distributions on $\varphi_{TY}$ and $|q^2|$
are expected also for $\eta\pi^-$ production in isospin exchanges 
with negative naturality, which means that the number of background events
obtained by fitting of Treiman-Yang angle distribution contains also
possible contribution from such exchanges.

Free parameters of the fit to the $q^2$-distribution (Fig. \ref{tcoul}) 
were number of events of Coulomb production of $\eta\pi^-$ (\ref{xprim})
and number of events of its hadronic production.
The interference of Coulomb and hadronic amplitudes was taken to be 
negligible, because the former is real, while the latter at high energies
is almost imaginary. 
For the slope parameter $b$ in (\ref{xdiffr}) the value of 
$b=50\mbox{ GeV}^{-2}$ was used, which corresponds to the 
diffractive production on $Be$ nuclei.
Varying this value by $\pm$10\% does not change the result.
The number of events of instrumental background was taken to be equal to 
that found from fitting the distribution on Treiman-Yang angle.
The effect of uncertainty in the number of background events 
on the error in determination of free parameters was taken into 
account by varying the number of background events within its standard
deviation.

The number of events in the Coulomb peak, in the range of 
$|q^2-q^2_{min}|<0.09\mbox{ GeV}^2$, is
\begin{equation}
N_{coul} = 109 \pm 23.
\end{equation}

In order to find the cross section, we used the results 
of \cite{bellini}\footnote{ Authors express 
their thanks to V.V.~Ezhela and Yu.I.~Ivanshin for the help in the 
extraction of necessary data from \protect\cite{bellini}.} ,
where the differential cross section of the reaction 
\begin{equation}
\pi^- Be \rightarrow \pi^+ \pi^- \pi^-Be
\label{diffr3pi}
\end{equation}
was measured at 40 GeV, 
and also the results of the VES experiment on reactions 
\begin{eqnarray}
\pi^- Be & \rightarrow & \pi^+ \pi^- \pi^-                 Be    \nonumber \\
\pi^- Be & \rightarrow & a_2^-(\pi^+ \pi^- \pi^-)            Be    \nonumber \\
\pi^- Be & \rightarrow & a_2^-(\eta(\pi^+ \pi^- \pi^0)\pi^-) Be.    \nonumber 
\end{eqnarray}
These data allow us to determine the cross section: 
\begin{equation}
\sigma_{coul} = 145 \pm 34 \mbox{ nb}.
\label{xsecexp}
\end{equation}

When calculating the error, we took into account the error in 
measurements of \cite{bellini} (5\%) 
and the systematic error (10\%) which corresponds to the uncertainty 
in the calculation of the efficiency of the VES setup.

Besides the process under study (\ref{box}), 
the Coulomb production of $a_2^-$-meson decaying into $\eta\pi^-$ 
also contributes to this cross section.
Assuming positive interference between amplitude (\ref{amplbox}) 
and amplitude of Coulomb production of $\eta\pi^-$ via $a_2^-$-meson, 
one can calculate the expected cross section of Coulomb production 
in the same range of $M_{\eta\pi^-}$ and $q^2$, using the experimentally 
measured \cite{a2ferbel} width of the decay $a_2^-(1320)\rightarrow\pi^-\gamma$ 
and parametrization of the dependence $F_{\eta\pi\pi\gamma}$ on $s,t,u$ 
from \cite{holstein}:
\begin{equation}
\sigma_{\eta\pi}^{th}=119\pm 13\mbox{ nb,}
\label{xtheor}
\end{equation}
(the quoted error corresponds to the errors of experimental measurements 
of widths of $a_2^-(1320)\rightarrow\pi\gamma$ 
and $\eta\rightarrow\pi^+\pi^-\gamma$ decays).
This value is in good agreement with the experimentally measured cross 
section (\ref{xsecexp}) and confirms strong dependence of 
$F_{\eta\pi\pi\gamma}$ on $t$.

For the description of the dependence of $F_{\eta\pi\pi\gamma}$ 
on kinematical variables in various theoretical models 
\cite{ko,holstein} the two substantial parameters are used ---
the value of $F_{\eta\pi\pi\gamma}$ in the chiral limit and the mass 
of vector meson in the $\pi\pi$ channel ($\rho$-meson).
Using the cross section measured in our experiment (\ref{xsecexp}) 
and the width of the decay $\eta\rightarrow\pi^+\pi^-\gamma$,
we calculated $F_{\eta\pi\pi\gamma}(0,0,0)$ and the effective mass of vector 
meson $\tilde m_{\rho}$ within the framework of the model \cite{holstein}.
We obtained the values 
\begin{equation}
F_{\eta \pi \pi \gamma}(0,0,0)=6.9\pm 0.7 \mbox{~GeV}^{-3}, \quad 
\quad \tilde m_{\rho}=900\pm 120 \mbox{~MeV}/c^2
\end{equation}
which agree respectively with (\ref{fsimple}) and the $\rho$-meson mass 
from \cite{pdg} within one standard deviation.

\section*{ Conclusion}

The cross section of the $\eta\pi^-$ production at low mass 
in the Coulomb field of beryllium nucleui 
at the beam momentum $p_{beam}=37\mbox{~GeV}/с$, 
$M_{\eta\pi^-}<1.18\mbox{~GeV}/c^2$ and $|q^2|<0.09\mbox{~GeV}^2$ 
was measured:
\begin{equation}
\sigma_{coul} = 145\pm 34\mbox{ nb}.
\nonumber
\end{equation}
This value is in good agreement with theoretical predictions 
\mbox{\cite{holstein,a2ferbel}}.

This work was supported by Russian Foundation for Basic Research 
(grant 98-02-16392) and INTAS-RFBR (grant 95-0267).

\newpage
\newsavebox{\acun}
\sbox{\acun}{\rotate {90} \hbox{N / 4 MeV/$c^2$} \endrotate}
\newsavebox{\ecun}
\sbox{\ecun}{\rotate {90} \hbox{N / 2.5 MeV/$c^2$} \endrotate}
\begin{figure}[ht]
\epsfxsize=\textwidth
\epsffile{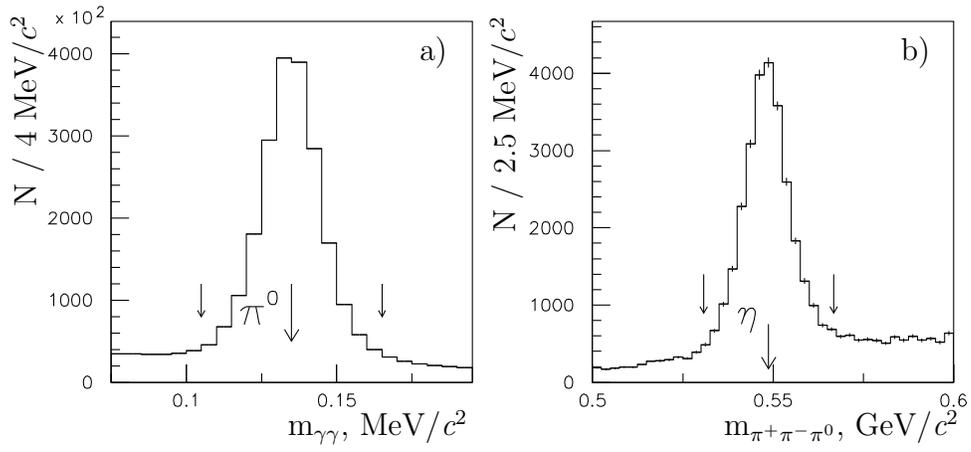}
\begin{picture}(160,0)
\put(47.5,22){m$_{\gamma\gamma}$, MeV/\it{c$^2$}}
\put(14,52.5){\usebox{\acun}}
\put(65,72){a)}
\put(105.5,22){m$_{\pi^+\pi^-\pi^0}$, GeV/\it{c$^2$}}
\put(77.7,49){\usebox{\ecun}}
\put(129,72){b)}
\end{picture}
\caption{Effective mass spectra of $\gamma\gamma$ (a) 
and $\pi^+\pi^-\pi^0$ (b).
The cuts used for the event selection are shown by arrows.}
\label{otbor}
\end{figure} 
\newpage
\newsavebox{\bcun}
\sbox{\bcun}{\rotate {90} \hbox{N / 40 MeV/$c^2$} \endrotate}
\begin{figure}[ht]
\epsfxsize=\textwidth
\epsfysize=0.7\textwidth
\epsffile{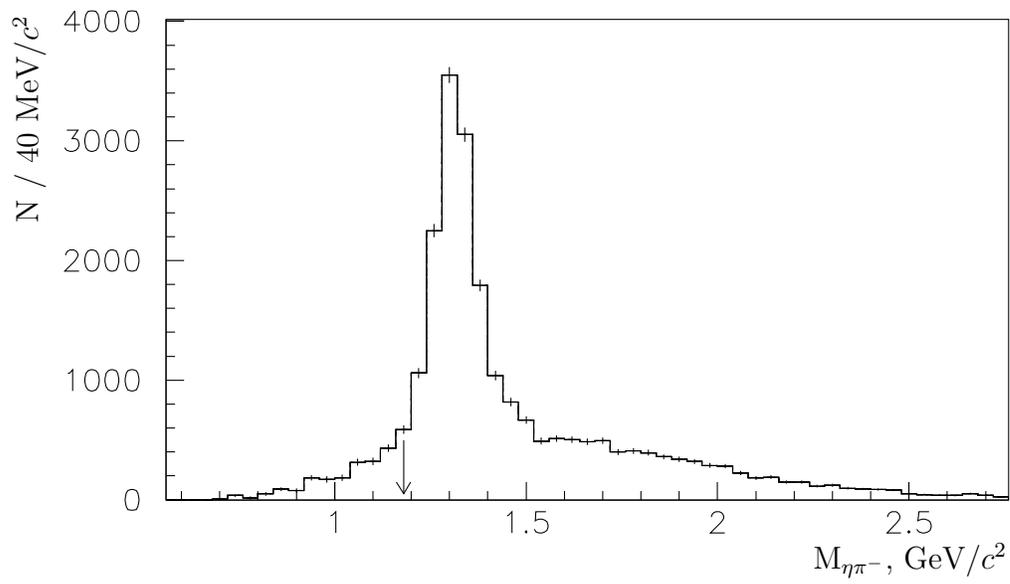}
\begin{picture}(160,0)
\put(110,20){M$_{\eta\pi^-}$, GeV/$c^2$}
\put(7,66){\usebox{\bcun}}
\end{picture}
\caption{The effective mass spectrum of $\eta\pi^-$. 
The arrow shows the cut used for the event selection 
$(M_{\eta\pi^-}<1.18\mbox{ GeV}/c^2)$.}
\label{m_etapi}
\end{figure}
\newpage
\newsavebox{\ccun}
\sbox{\ccun}{\rotate {90} \hbox{N / 0.003 GeV$^2$} \endrotate}
\begin{figure}[ht]
\epsfxsize=\textwidth
\epsffile{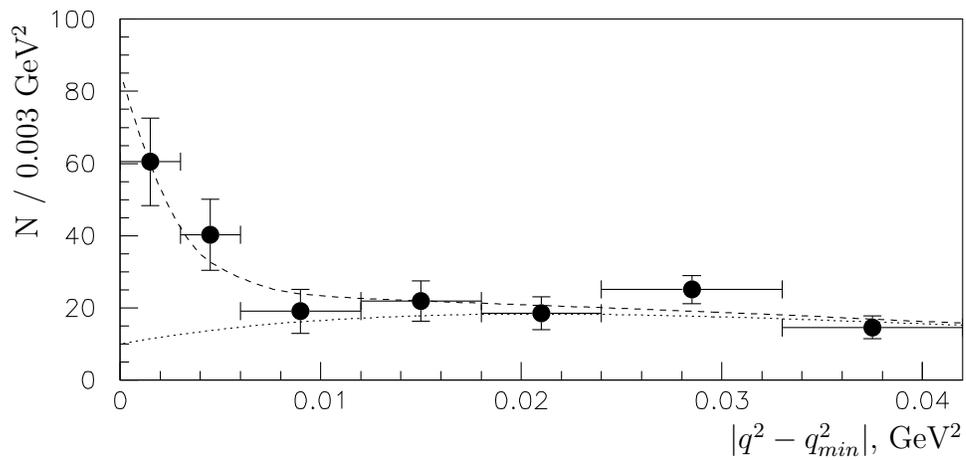}
\begin{picture}(160,0)
\put(105,20){$|q^2-q^2_{min}|$, GeV$^2$}
\put(13,48){\usebox{\ccun}}
\end{picture}
\caption{The $|q^2-q^2_{min}|$ distribution for events with 
$m_{\eta\pi^-}<1.18\mbox{ GeV}/c^2$.
The dashed line is the result of the fit (see text); 
the dotted line shows the contribution of hadronic production 
of $\eta\pi^-$ and background.}
\label{tcoul}
\end{figure}
\newpage
\newsavebox{\dcun}
\sbox{\dcun}{\rotate {90} \hbox{Events} \endrotate}
\begin{figure}[ht]
\epsfxsize=\textwidth
\epsffile{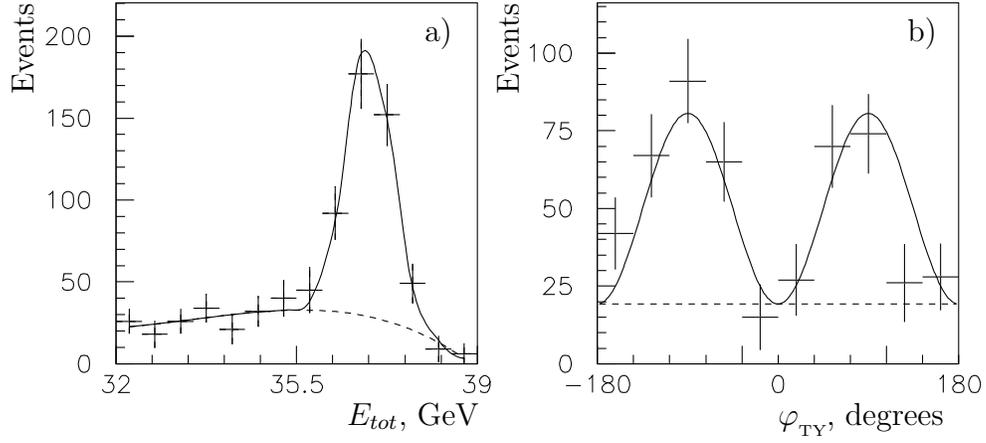}
\begin{picture}(160,0)
\put(54.7,21){$E_{tot}$, GeV}
\put(112,21){$\varphi$\raisebox{-1.5mm}{\tiny TY}, degrees}
\put(65,72){a)}
\put(129.2,72){b)}
\put(13,65){\usebox{\dcun}}
\put(77.5,65){\usebox{\dcun}}
\end{picture}
\caption{Distributions on total energy (a) and Treiman-Yang angle (b)
for the $\eta\pi^-$ events with $M_{\eta\pi^-}<1.18\mbox{ GeV}$, 
$0.01<|q^2-q^2_{min}|<0.1\mbox{ GeV}^2$.}
\label{figfi}
\end{figure}
\end{document}